\documentclass[a4paper,showkeys, preprint,13pt]{revtex4}  %showkeys, onecolumn, prl,showpacs,twocolumn,
\usepackage{latexsym}
\usepackage[dvipdfm,colorlinks=true,linkcolor=blue,anchorcolor=blue,citecolor=blue]{hyperref}
\usepackage[hmargin={1cm,1cm}, vmargin={2cm,2cm}]{geometry}
\usepackage{float}
\usepackage{graphicx,amsfonts,amsmath,amssymb,amsthm,amscd}
\usepackage{epsfig}
\usepackage{bm}
\usepackage{xr}
\usepackage{indentfirst}
\usepackage{appendix}
\usepackage{changepage}

\newcommand{\ds}{\mathrm{d}s}
\newcommand{\dt}{\mathrm{d}t}

\date{\today}

\begin{document}

\title{\bf Entropy production related properties of first passage process}

\author{Yunxin Zhang} \email[Email: ]{xyz@fudan.edu.cn}
\affiliation{Laboratory of Mathematics for Nonlinear Science, Shanghai Key Laboratory for Contemporary Applied Mathematics, Centre for Computational Systems Biology, School of Mathematical Sciences, Fudan University, Shanghai 200433, China.}

\begin{abstract}
With nontrivial entropy production, first passage process is one of the most common nonequilibrium process in stochastic thermodynamics. Using one dimensional birth and death precess as a model framework, approximated expressions of mean first passage time (FPT), mean total number of jumps (TNJ), and their coefficients of variation (CV), are obtained for the case far from equilibrium. Consequently, uncertainty relations for FPT and TNJ are presented. Generally, mean FPT decreases exponentially with entropy production, while mean TNJ decreases exponentially first and then tends to a starting site dependent limit. For forward biased process, the CV of TNJ decreases exponentially with entropy production, while that of FPT decreases exponentially first and then tends to a starting site dependent limit. For backward biased process, both CVs of FPT and TNJ tend to one for large absolute values of entropy production. Related properties about the case of equilibrium are also addressed briefly for comparison.
\end{abstract}


%\pacs{111,eee}

\maketitle

\section{Introduction}
%\paragraph*{Introduction}
Entropy production, which characterizes the degree of irreversibility (or dissipation), plays a key role in nonequilibrium stochastic thermodynamics \cite{Sekimoto2010,Seifert2012Stochastic,Shiraishi2019,Seifert2019}. In recent decades, many studies have been devoted to optimize the performance of stochastic thermodynamic machines, usually by reducing the total entropy production  \cite{Schmiedl2007,Gomez2008,Aurell2011,Golubeva2012Efficiency,Muratore2013,Horowitz2018,Zhang20191,Zhang2020,Abiuso2020,Sahae2021,Remlein2021,Gronchi2021}. A general relation between nonequilibrium currents and entropy production, the thermodynamic uncertainty relation (TUR), is discovered and serves as a fundamental principle of nonequilibrium thermodynamics \cite{Barato2015,Proesmans2017,Dechant2018,Macieszczak2018,Timpanaro2019,Hasegawa2019,Ito2020,Liu2020,Horowitz2020,Vu2020,Koyuk2020,Hasegawa2020, Falasco2020,Dechant2020}, which dictates that the precision of a nonequilibrium time-integrated current observable $J$ is bounded from below by the inverse of the total entropy production (EP) $\sigma$, $(\langle J^2\rangle-\langle J\rangle^2)/\langle J\rangle^2 \ge 2/\langle\sigma\rangle$, with $\langle \cdot\rangle$ the average of random variables.

Recently, uncertainty relation between FPT and EP, sometimes called kinetic uncertainty relation (KUR), has attracted much attention. In \cite{Garrahan2017}, uncertainty relation for FPT is obtained by large deviation principle, which states that the precision of estimation of the FPT is
limited by the total average dynamical activity. In \cite{Gingrich2017}, bounds on FPT fluctuation for currents are obtained by transferring bounds on the large-deviation function for currents, which show the FPT fluctuation is also bound by dissipation. In \cite{Falasco20201}, a dissipation-time uncertainty relation is presented, which shows that the EP rate bounds the rate at which physical processes can be performed in stochastic systems far from equilibrium. In \cite{Hiura2021}, the KUR on FPT is derived from the information inequality at stopping times, which shows that the precision of FPT is bounded from above by the mean TNJ. In
\cite{Kuznets-Specke2021}, transition between long-lived states is discussed, and they found that there is a basic speed limit relating the typical excess heat dissipated throughout a transition and the rate amplification achievable. In \cite{Pal2021}, a TUR for FPT on continuous time Markov chains is derived, which shows
that the coefficient of variation of the FPT is bounded from below by an expression that combines
the entropy production from the transitions between internal states and the fluxes to absorbing boundaries.

\begin{figure}[htbp]
	\includegraphics[width=0.8\linewidth]{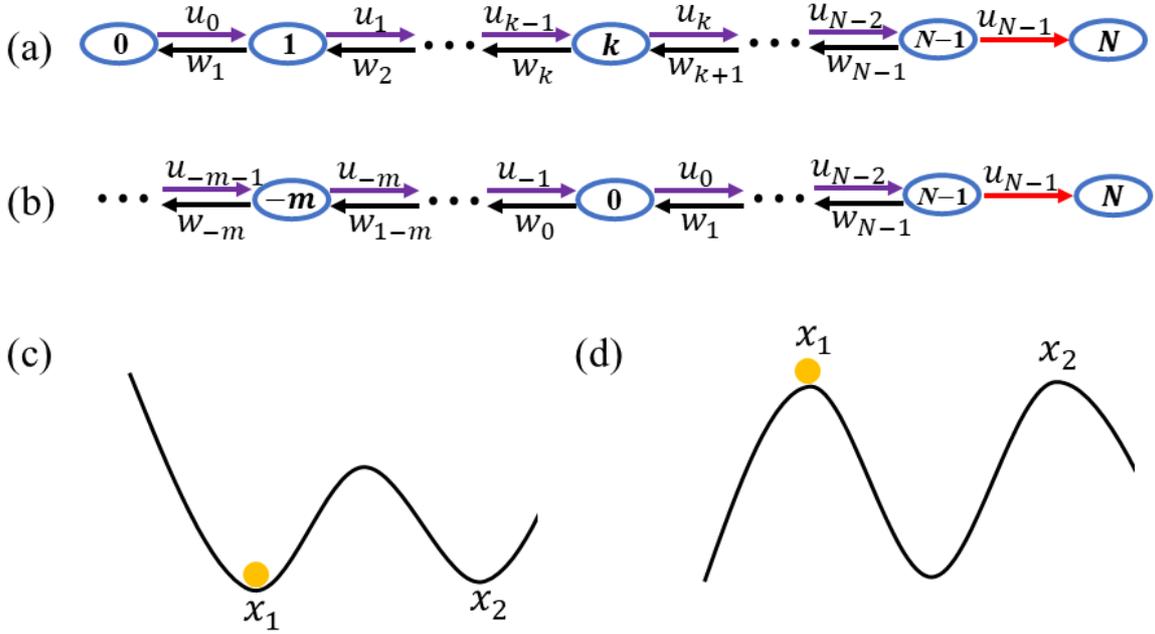}
	\caption{First passage process discussed in this study, with site $N$ an absorbing boundary. \textbf{(a)} Finite state cases with site $0$ a reflecting boundary. \textbf{(b)} Infinite state cases with $-\infty$ a reflecting boundary. \textbf{(c,d)} Two typical examples of first passage process, with the random particle starting from position $x_1$ to reach absorbing boundary $x_2$.  } \label{FigSchematic}
\end{figure}

In general, EP (or heat dissipation) is essential to first passage process. To find out EP related properties, this study considers a simple one-dimensional birth and death process (continuous-time Markov chain as depicted in Fig.~\ref{FigSchematic}\textbf{(a)}), where the forward jump rate from site $k$ to $k+1$ is denoted by $u_k$, and the backward jump rate from site $k$ to $k-1$ is denoted by $w_k$. The boundary site $N$ is assumed to be an absorbing boundary, while the boundary site $0$ is assumed to be a reflecting boundary. Results are similar if boundary site 0 is also an absorbing boundary, or if there are infinite sites as depicted in Fig.~\ref{FigSchematic}\textbf{(b)}.
This model process can be used sometimes to discuss the transition in a system with two stable steady states as shown in Fig.~\ref{FigSchematic}\textbf{(c,d)} \cite{Kampen2007,Gardiner2010}.%,Kuznets-Specke2021}.

\section{First passage process with finite states}
The probability density $p_k(s,t)$ that a random particle initiated at site $k$ to reach the absorbing boundary $N$ at time $t$ and with EP $s$ satisfies the following backward master equation
\begin{equation}\label{Eqf}
\partial_t p_k(s,t)=u_kp_{k+1}(s-\ln(u_k/w_{k+1}),t)+w_kp_{k-1}(s-\ln(w_k/u_{k-1}),t)-(u_k+w_k)p_k(s,t),
\end{equation}
with $\ln(u_k/w_{k+1})$ and $\ln(w_k/u_{k-1})$ entropy productions ($k_B=1$) during one forward and backward jump, respectively \cite{Seifert2005,Seifert2012Stochastic,Teza2020}. From Eq.~(\ref{Eqf}), the mean EP $S_k:=\int_{-\infty}^{\infty}s\int_0^{\infty}p_k(s,t)\dt\ds$ during the passage process from site $k$ to boundary $N$ satisfies
\begin{equation}\label{EqS}
u_kS_{k+1}+w_kS_{k-1}-(u_k+w_k)S_k=-\left(u_k\ln\frac{u_k}{w_{k+1}}+w_k\ln\frac{w_k}{u_{k-1}}\right).
\end{equation}
The mean FPT $T_k:=\int_0^{\infty}t\int_{-\infty}^{\infty}p_k(s,t)\ds\dt$, and the second moment $T_k^{(2)}:=\int_0^{\infty}t^2\int_{-\infty}^{\infty}p_k(s,t)\ds\dt$, satisfy
%\begin{subequations}\label{EqTandT2}
\begin{eqnarray}
&&u_kT_{k+1}+w_kT_{k-1}-(u_k+w_k)T_k=-1,\label{EqT}\\
&&u_kT_{k+1}^{(2)}+w_kT_{k-1}^{(2)}-(u_k+w_k)T_k^{(2)}=-2T_k.\label{EqT2}
\end{eqnarray}
%\end{subequations}

The probability $q_k(n)$ that a random particle initiated at site $k$ to reach the absorbing boundary $N$ with TNJ $n$ satisfies the following backward master equation
\begin{equation}\label{Eqq}
u_kq_{k+1}(n-1)+w_kq_{k-1}(n-1)-(u_k+w_k)q_k(n)=0.
\end{equation}
So, the mean TNJ $n_k:=\sum_{n=0}^{\infty}nq_k(n)$, and the second moment $n_k^{(2)}:=\sum_{n=0}^{\infty}n^2q_k(n)$, %before reaching the absorbing boundary $N$,
satisfy
%\begin{subequations}\label{Eqnandn2}
\begin{eqnarray}
&&u_kn_{k+1}+w_kn_{k-1}-(u_k+w_k)n_k=-(u_k+w_k),\label{Eqnk}\\
&&u_kn_{k+1}^{(2)}+w_kn_{k-1}^{(2)}-(u_k+w_k)n_k^{(2)}=-(2n_k-1)(u_k+w_k).\label{Eqnk2}
\end{eqnarray}
%\end{subequations}

By usual algebraic calculations, explicit expressions of mean EP $S_k$, mean FPT $T_k$, mean TNJ $n_k$, as well as the second moments $T_k^{(2)}$ and $n_k^{(2)}$ can be obtained by solving Eqs.~(\ref{EqS},\ref{EqT},\ref{EqT2},\ref{Eqnk},\ref{Eqnk2}), see \textbf{Supplemental Materials}.
To illustrate the influence of EP $S_k$ on FPT and TNJ, we consider the particular case with $u_k\equiv u$ and $w_k\equiv w=ru$. It can be shown that,
\begin{eqnarray}
%\begin{aligned}
S_k(r)&=&-(N-k-1)\ln r, \label{EqSTspecialOne1}\\
T_k(r)&=&\frac{N-k}{u(1-r)}+\frac{r^{N+1}-r^{k+1}}{u(1-r)^2},\label{EqSTspecialOne2}\\
n_k(r)&=&\frac{(1+r)(N-k)}{1-r}+\frac{(1+r)(r^{N+1}-r^{k+1})}{(1-r)^2},\label{EqSTspecialOne4}
\end{eqnarray}
and
\begin{eqnarray}
%\begin{aligned}
T_k^{(2)}(r)-T_k^{2}(r)&=&\frac{(N-k)(1+r)}{u^2(1-r)^3}+\frac{r^{2N+2}-r^{2k+2}}{u^2(1-r)^4}+o(r),\label{EqSTspecialOne6}\\
n_k^{(2)}(r)-n_k^{2}(r)&=&\frac{4(N-k)r(1+r)}{(1-r)^3}+\frac{(1+r)^2(r^{2N+2}-r^{2k+2})}{(1-r)^4}+o(r),\label{EqSTspecialOne7}
%\end{aligned}
\end{eqnarray}
with $o(r)$ the high (low) order term for $r\to0$ ($r\to\infty$).

\begin{figure}[htbp]
\includegraphics[width=0.5\linewidth]{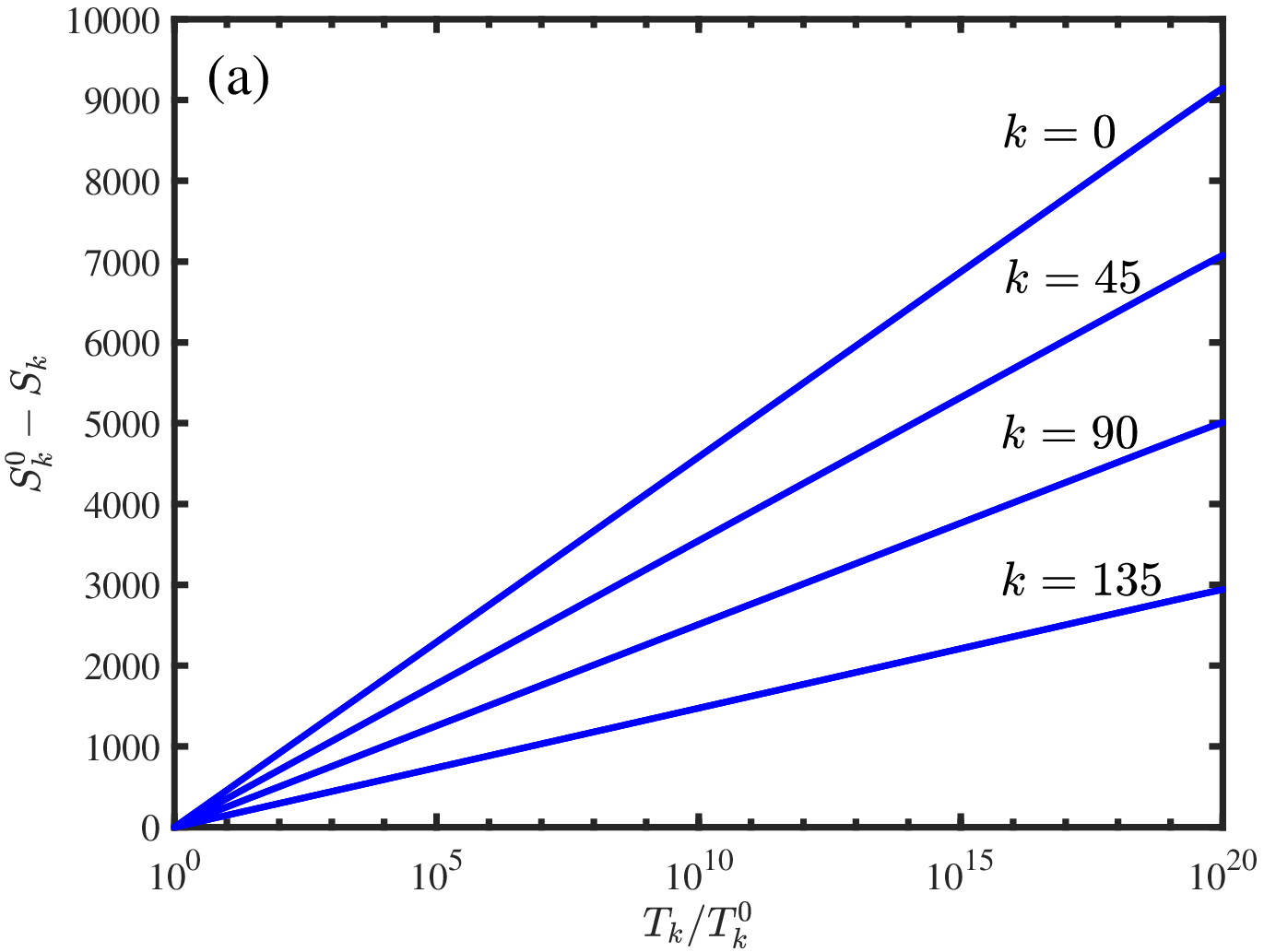}\includegraphics[width=0.5\linewidth]{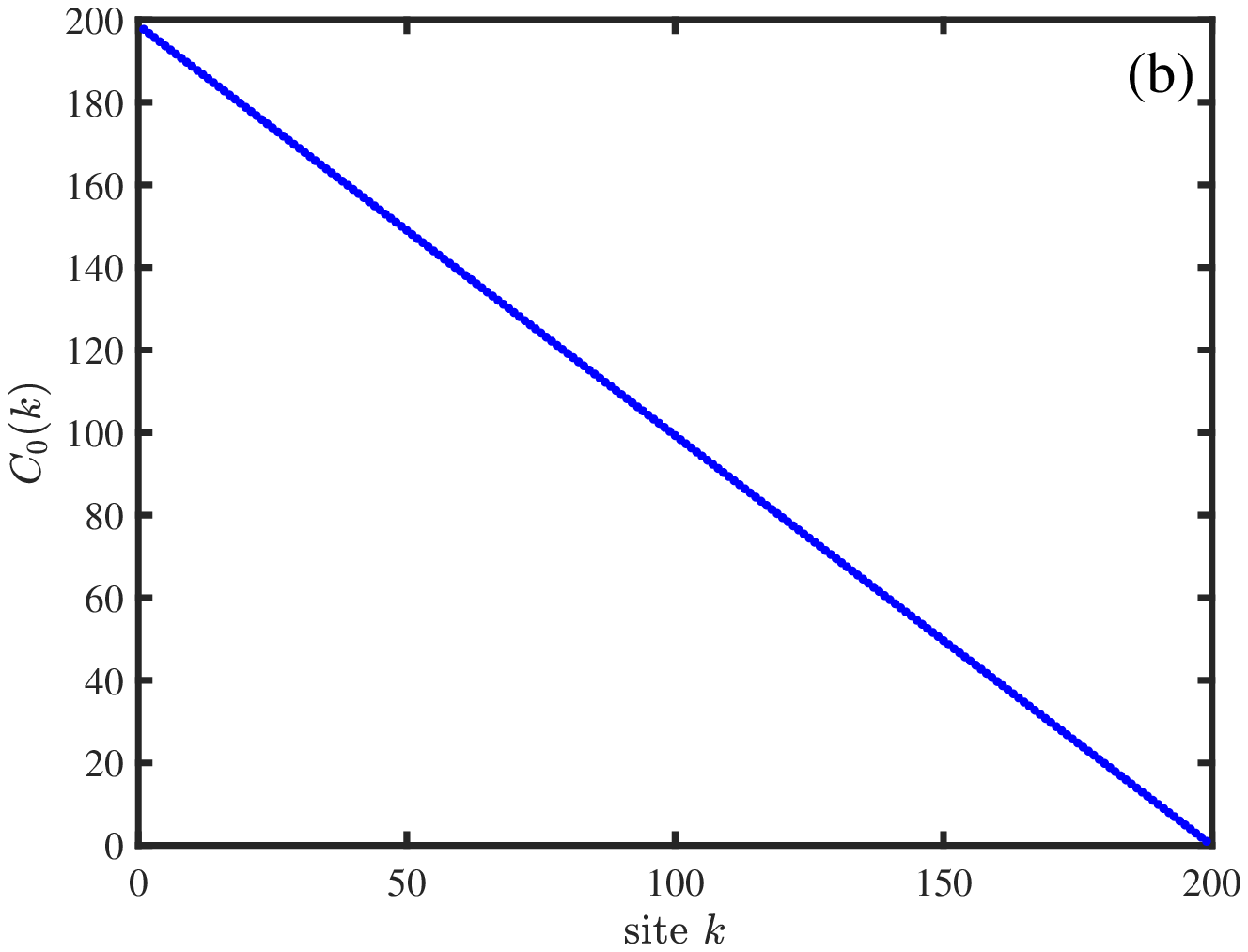}
	\caption{Relationship between EP and the ratio of mean FPT for the process depicted in Fig.~\ref{FigSchematic}\textbf{(a)}, with $N=200$ therein. For each starting site $k$, 1000 values of $S_k$ and $T_k$ are calculated, see Eqs.~(\ref{EqSSsolutionOne},\ref{EqSTsolutionOne}). In the first calculation, all forward rates $\{u_k^{(1)}\}$ and backward rates $\{w_k^{(1)}\}$ are randomly selected in interval $(0.5, 3)$. Then in following calculations, backward rates $\{w_k\}$ remain unchanged while forward rates change according to $u_k^{(l+1)}=\gamma u_k^{(l)}$ with $\gamma=1.05$. For convenience, $T_k^0$ is chosen to be the minimal value of $T_k$, and $S_k^0$ is the EP according to $T_k^0$. \textbf{(a)} $S^0_{k}-S_{k}$ versus $T_{k}/T^0_{k}$, with starting site $k=0, 45, 90, 135$ respectively. Results show that $T_k/T_k^0\approx C_1(k)\exp(-(S_k-S_k^0)/C_0(k))$. \textbf{(b)} $C_0(k)$ obtained by fitting to $T_k/T_k^0=C_1(k)\exp(-(S_k-S_k^0)/C_0(k))$, see Eq.~(\ref{EqLimitCase1Relation}). For this example $C_0(k)\approx N-k-1$ while $C_1(k)\approx0.9$ is almost constant. }\label{FigCase1OneAbsorb}
\end{figure}

\subsection{Forward biased process with $r=w/u<1$}\label{SectionRsmall1}
If the backward rate $w$ is fixed while forward rate $u$ is large enough, {\it i.e.}, $r\ll1$, then Eq.~(\ref{EqSTspecialOne2}) implies $T_k\sim{(N-k)r}/{w}$, and
\begin{eqnarray}\label{EqLimitCase1Relation}
\frac{T_k(r_1)}{T_k(r_0)}\sim \frac{r_1}{r_0}=\exp\left(-\frac{{S_k(r_1)}-{S_k(r_0)}}{N-k-1}\right)=:\exp\left(-\frac{\Delta S_k}{N-k-1}\right).
\end{eqnarray}
Numerical calculations show this relationship between ${T_k(r_1)}/{T_k(r_0)}$ and $\Delta S_k$ is still true for general nonconstant rates $\{u_k\}$ and $\{w_k\}$ which satisfy $w_k/u_k\equiv r\ll1$, see Fig.~\ref{FigCase1OneAbsorb}.

\begin{figure}[htbp]
\includegraphics[width=0.8\linewidth]{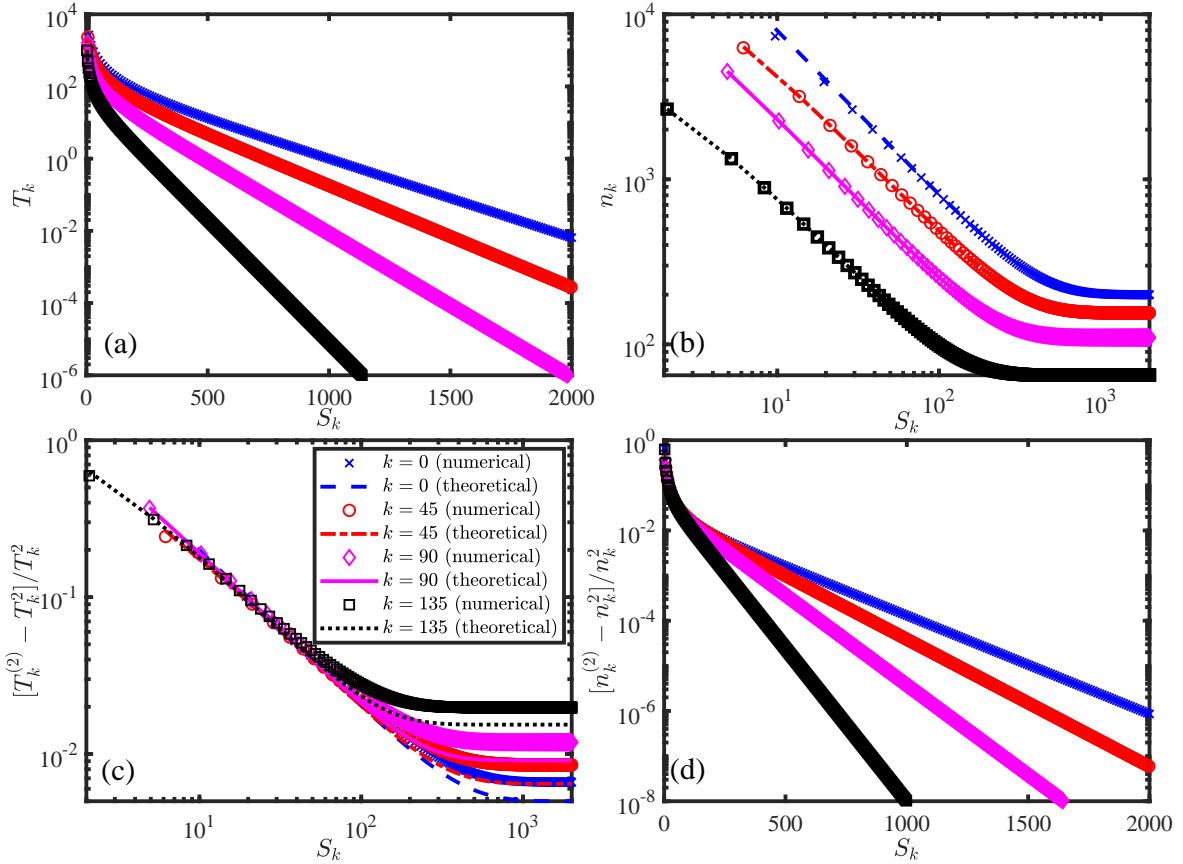}
	\caption{Mean FPT $T_k$, mean TNJ $n_k$, and $[T_k^{(2)}-T_k^{2}]/T_k^2$, $[n_k^{(2)}-n_k^{2}]/n_k^2$, with the change of EP $S_k$, and for starting site $k=0,45,90,135$ respectively. Points are obtained numerically by formulations given in Eqs.~(\ref{EqSSsolutionOne},\ref{EqSTsolutionOne},\ref{EqST2solutionOne},\ref{EqSnksolutionOne},\ref{EqSnk2solutionOne}), and with parameter values randomly selected in the same intervals as in Fig.~\ref{FigCase1OneAbsorb}. Lines are obtained from the bounds given in Eqs.~(\ref{EqLimitCase1Uncertainty1},\ref{EqLimitCase1Uncertainty4},\ref{EqLimitCase1Uncertainty3},\ref{EqLimitCase1Uncertainty6}), with $r$ replaced by $\exp(-S_k/(N-k-1))$, and $w$ in Eq.~(\ref{EqLimitCase1Uncertainty1}) replaced by the harmonic mean of $\{w_i\}_{i=k}^{N-1}$.
}\label{FigCase1OneAbsorbTNVar}
\end{figure}

From Eqs.~(\ref{EqSTspecialOne2},\ref{EqSTspecialOne4},\ref{EqSTspecialOne6},\ref{EqSTspecialOne7}), for $r\ll1$,
\begin{eqnarray}  %\alt
&&T_k(r)\alt\frac{(N-k)r}{w(1-r)},\label{EqLimitCase1Uncertainty1}\\
&&n_k(r)\alt \frac{(1+r)(N-k)}{(1-r)},\label{EqLimitCase1Uncertainty4}\\
&&T_k^{(2)}(r)-T_k^{2}(r)\sim\frac{(N-k)(1+r)r^2}{w^2(1-r)^3},\label{EqLimitCase1Uncertainty2}\\
&&n_k^{(2)}(r)-n_k^{2}(r)\sim\frac{4(N-k)r(1+r)}{(1-r)^3}.\label{EqLimitCase1Uncertainty2-3}
\end{eqnarray}
Therefore,
\begin{eqnarray}
%\begin{aligned}
\textrm{CV}[T_k(r)]:=\frac{T_k^{(2)}(r)-T_k^{2}(r)}{T_k^{2}(r)}\agt \frac{1+r}{(N-k)(1-r)}
=\frac{1+\exp(-S_k(r)/(N-k-1))}{(N-k)[1-\exp(-S_k(r)/(N-k-1))]}\to\frac{1}{N-k},\label{EqLimitCase1Uncertainty3}
%\end{aligned}
\end{eqnarray}
which means that CV of FPT decreases with EP $S_k$, and tends to its lower limit $\textrm{CV}[T_k(0)]=1/(N-k)$ with $S_k\to\infty$, see Fig~\ref{FigCase1OneAbsorbTNVar}. Eq.~(\ref{EqLimitCase1Uncertainty3}) can be rewritten as follows,
\begin{eqnarray}
%\begin{aligned}
\frac{T_k^{2}(r)}{T_k^{(2)}(r)-T_k^{2}(r)}\alt \frac{(N-k)(1-r)}{1+r}\leq n_k(r),\label{EqLimitCase1Uncertainty5}
%\end{aligned}
\end{eqnarray}
which is the KUR obtained in \cite{Hiura2021}. From Eq.~(\ref{EqLimitCase1Uncertainty3}) and Fig~\ref{FigCase1OneAbsorbTNVar}, $\textrm{CV}[T_k(r)]-\textrm{CV}[T_k(0)]\agt 2r/(N-k)$, so CV of FPT decreases exponentially to its limit $\textrm{CV}[T_k(0)]$ when $S_k\to\infty$.

For TNJ, we obtain from Eqs.~(\ref{EqLimitCase1Uncertainty4},\ref{EqLimitCase1Uncertainty2-3}) that
\begin{eqnarray}
%\begin{aligned}
\textrm{CV}[n_k(r)]:&=&\frac{n_k^{(2)}(r)-n_k^{2}(r)}{n_k^{2}(r)}\agt \frac{4r}{(N-k)(1-r^2)}\cr
&=&\frac{4\exp(-S_k(r)/(N-k-1))}{(N-k)[1-\exp(-2S_k(r)/(N-k-1))]}\sim \frac{4\exp(-S_k(r)/(N-k-1))}{(N-k)},\label{EqLimitCase1Uncertainty6}
%\end{aligned}
\end{eqnarray}
which decreases to 0 exponentially with entropy $S_k$, see Fig~\ref{FigCase1OneAbsorbTNVar}.

Eqs.~(\ref{EqLimitCase1Uncertainty1},\ref{EqLimitCase1Uncertainty4}) and Fig~\ref{FigCase1OneAbsorbTNVar} show that, with $S_k\to\infty$, $T_k$ decreases to 0 and $n_k$ decreases to its lower limit $N-k$, both exponentially.

\begin{figure}[htbp]
\includegraphics[width=0.8\linewidth]{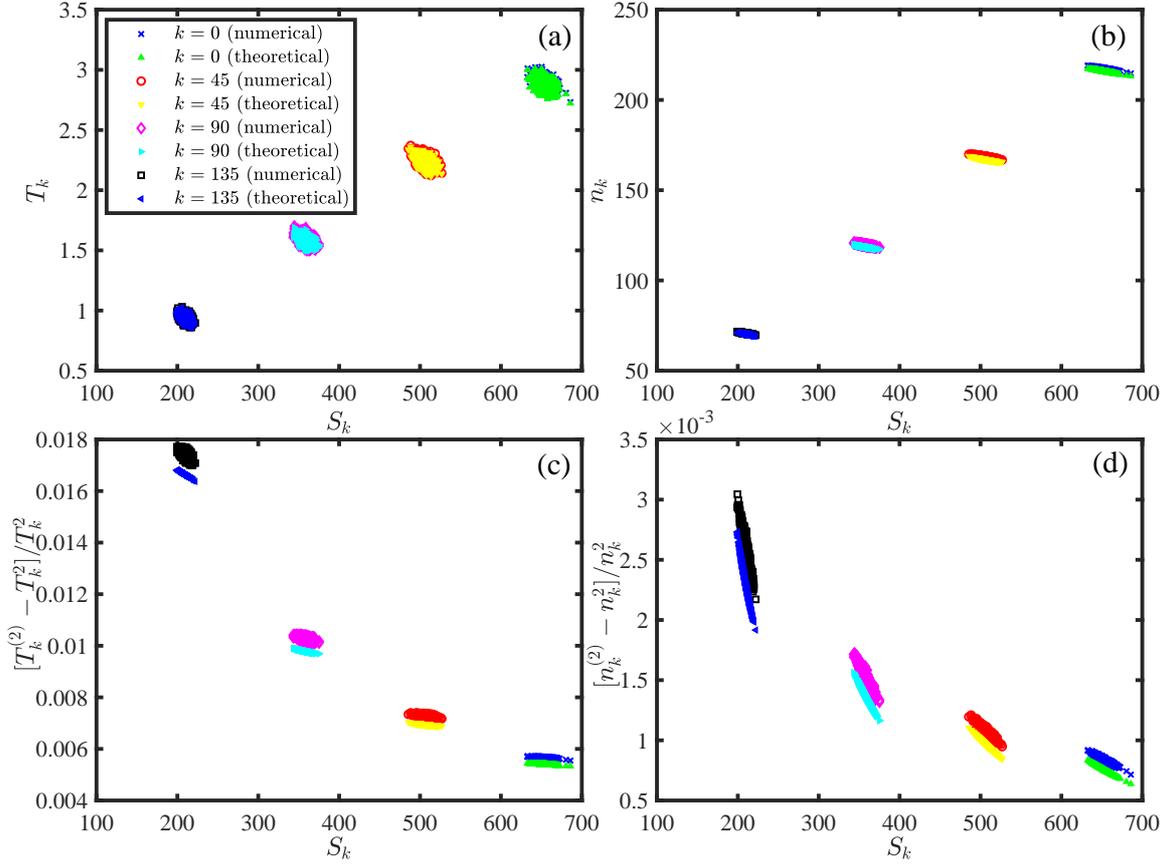}
	\caption{Results for randomly selected values of both forward rates $\{u_i\}$ and backward rates $\{w_i\}$. We do 1000 calculations, and in each calculation, $N=200$ is fixed, while $\{w_i\}$ are randomly selected in interval $(1,5)$, and $\{u_i\}$ are randomly selected in interval $(50,100)$. The theoretical points (marked by varies triangles) are obtained according to the bounds given in Eqs.~(\ref{EqLimitCase1Uncertainty1},\ref{EqLimitCase1Uncertainty4},\ref{EqLimitCase1Uncertainty3},\ref{EqLimitCase1Uncertainty6}), with $r$ replaced by $\exp(-S_k/(N-k-1))$, and $w$ in Eq.~(\ref{EqLimitCase1Uncertainty1}) replaced by the harmonic mean of $\{w_i\}_{i=k}^{N-1}$. Numerical values are obtained as in Fig.~\ref{FigCase1OneAbsorbTNVar}.
}\label{FigRandominRateValuesUlarge}
\end{figure}

On the other hand, if forward rate $u$ is fixed while backward rate $w\to0$, then
Eq.~(\ref{EqSTspecialOne2}) gives $T_k(r)\alt  {(N-k)}/{[u(1-r)]}$, and results in Eq.~(\ref{EqLimitCase1Uncertainty4},\ref{EqLimitCase1Uncertainty3},\ref{EqLimitCase1Uncertainty5},\ref{EqLimitCase1Uncertainty6}) still hold true, see Fig.~\ref{FigCase3OneAbsorbTNVar}.
Since $T_k(r)-T_k(0)\sim(N-k)r/u$, $T_k(r)$ decreases exponentially  to its limit $(N-k)/u$ with $S_k\to\infty$.

Results given in Eqs.~(\ref{EqLimitCase1Uncertainty1},\ref{EqLimitCase1Uncertainty4},\ref{EqLimitCase1Uncertainty3},\ref{EqLimitCase1Uncertainty6})
are also true for nonconstant rates $\{u_k\}$ and $\{w_k\}$ which satisfy $w_k/u_k\equiv r\ll1$ (or $S_k\gg0$ equivalently), see Fig~\ref{FigCase1OneAbsorbTNVar}. Generally, if both forward rates $\{u_k\}$ and backward rates $\{w_k\}$ are chosen randomly but keeping EP $S_k$ positive, inequalities given in Eqs.~(\ref{EqLimitCase1Uncertainty3},\ref{EqLimitCase1Uncertainty6}) are also true, but the ones given in Eqs.~(\ref{EqLimitCase1Uncertainty1},\ref{EqLimitCase1Uncertainty4}) may not, see Fig~\ref{FigRandominRateValuesUlarge}. However, for large $S_k$, mean FPT $T_k$ and mean TNJ $n_k$ can be well approximated by the upper bounds given in Eqs.~(\ref{EqLimitCase1Uncertainty1},\ref{EqLimitCase1Uncertainty4}).  Note that, for randomly selected rates $\{u_k\}$ and $\{w_k\}$, the ratio $r$ in Eqs.~(\ref{EqLimitCase1Uncertainty1},\ref{EqLimitCase1Uncertainty4},\ref{EqLimitCase1Uncertainty3},\ref{EqLimitCase1Uncertainty6})
should be replaced by $\exp(-S_k/(N-k-1))$, and $w$ (or $u$) in Eq.~(\ref{EqLimitCase1Uncertainty1}) should be replaced by the harmonic mean of $\{w_i\}_{i=k}^{N-1}$ (or $\{u_i\}_{i=k}^{N-1}$).

\subsection{Backward biased process with $r=w/u>1$}\label{SectionRlarge1}
If backward rate $w$ is fixed while forward rate $u$ is small enough ($r\gg1$), then Eqs.~(\ref{EqSTspecialOne2}) implies
$T_k(r)\sim{r^N}/{w}$. So,
\begin{eqnarray}
\frac{T_k(r_1)}{T_k(r_0)}&\sim&\left(\frac{r_1}{r_0}\right)^N=\exp\left(-\frac{N\Delta S_k}{N-k-1}\right). \label{EqLimitCase2Relation-1}
\end{eqnarray}
Again this approximation is still true for nonconstant rates $\{u_k\}$ and $\{w_k\}$ which satisfy $w_k/u_k\equiv r\gg1$, see Fig.~\ref{FigCase2OneAbsorb}.

According to Eqs.~(\ref{EqSTspecialOne2},\ref{EqSTspecialOne4},\ref{EqSTspecialOne6},\ref{EqSTspecialOne7}), for $r\gg1$,
\begin{eqnarray}
&&T_k(r)\alt  \frac{r^{N+2}-r^{k+2}}{w(1-r)^2},\quad
n_k(r)\alt \frac{(1+r)(r^{N+1}-r^{k+1})}{(1-r)^2},\label{EqLimitCase2Uncertainty1}\\
&&T_k^{(2)}(r)-T_k^{2}(r)\sim\frac{r^{2N+4}-r^{2k+4}}{w^2(1-r)^4}.\label{EqLimitCase2Uncertainty2}\\
&&n_k^{(2)}(r)-n_k^{2}(r)\sim\frac{(1+r)^2 (r^{2N+2}-r^{2k+2})}{(1-r)^4}.\label{EqLimitCase2Uncertainty2-3}
\end{eqnarray}
Therefore,
\begin{eqnarray}
\textrm{CV}[T_k(r)]\sim \textrm{CV}[n_k(r)]
\substack{>\\\sim} \frac{r^{N-k}+1}{r^{N-k}-1}
=\frac{\exp(-(N-k)S_k(r)/(N-k-1))+1}{\exp(-(N-k)S_k(r)/(N-k-1))-1}\to1.\label{EqLimitCase2Uncertainty3}
\end{eqnarray}

Similar as in the {\it forward biased} cases, % Sec.~\ref{SectionRsmall1},
results given in Eqs.~(\ref{EqLimitCase2Uncertainty1},\ref{EqLimitCase2Uncertainty3}) are still valid for nonconstant rates $\{u_k\}$ and $\{w_k\}$ which satisfy $w_k/u_k\equiv r\gg1$, see Fig.~\ref{FigCase2OneAbsorbTNVar}. Generally, if both  $\{u_k\}$ and $\{w_k\}$ are selected randomly but keeping EP $S_k$ negative, then for the absolute value of $S_k$ large enough, {\it i.e.}, for process far from equilibrium, $\textrm{CV}[T_k]$ and $\textrm{CV}[n_k]$ are almost one, and $T_k, n_k$ can be well approximated by the bounds given in Eq.~(\ref{EqLimitCase2Uncertainty1}), see Fig.~\ref{FigRandominRateValuesUsmall}. It can be easily shown that the KUR given in Eq.~(\ref{EqLimitCase1Uncertainty5}) is also valid.

Meanwhile, if forward rate $u$ is fixed while backward rate $w=ru$ tends to infinity, then Eq.~(\ref{EqSTspecialOne2}) implies
$T_k(r)\sim{r^{N-1}}/{u}$, which means
\begin{eqnarray}
\frac{T_k(r_1)}{T_k(r_0)}&\sim&\left(\frac{r_1}{r_0}\right)^{N-1}=\exp\left(-\frac{(N-1)\Delta S_k}{N-k-1}\right). \label{EqLimitCase4Relation-1}
\end{eqnarray}
At the same time, $T_k(r)\alt  {(r^{N+1}-r^{k+1})}/{[u(1-r)^2]}$, and results for $n_k(r)$ and $\textrm{CV}[T_k(r)], \textrm{CV}[n_k(r)]$ given in  Eqs.~(\ref{EqLimitCase2Uncertainty1},\ref{EqLimitCase2Uncertainty3}) still hold, see Fig.~\ref{FigCase4OneAbsorbTNVar}.

In summary, ${T_k(r_1)}/{T_k(r_0)}\sim C_1(k)\exp(-{\Delta S_k}/C_0(k))$ if backward rate $\{w_k\}$ are fixed while forward rate $\{u_k\}$ tend to infinity or zero, or forward rate $\{u_k\}$ are fixed while backward rate $\{w_k\}$ tend to infinity, see Eqs.~(\ref{EqLimitCase1Relation},\ref{EqLimitCase2Relation-1},\ref{EqLimitCase4Relation-1}) and Figs.~\ref{FigCase1OneAbsorb},\ref{FigCase2OneAbsorb},\ref{FigCase4OneAbsorbTNVar}. The ratio $r$ of backward rate to forward rate tends to 0 or infinity means the process is far from equilibrium. So the results obtained here are consistent with the ones obtained previously \cite{Kuznets-Specke2021}.
For fixed $\{u_k\}$ , if backward rates $\{w_k\}$ tend to zero, then the process becomes a deterministic motion. So $T_k\to1/u_k+\cdots+1/u_{N-1}=(N-k)/u$ and $n_k\to N-k$, see Eqs.~(\ref{EqSTspecialOne2},\ref{EqSTspecialOne4}) and Fig.~\ref{FigCase3OneAbsorbTNVar}. For either $\{u_k\}$ tend to infinity or $\{w_k\}$ tend to 0, mean TNJ $n_k$ decreases to its lower bound $N-k$, $\textrm{CV}[T_k]$ decreases to its lower bound $1/(N-k)$, and $\textrm{CV}[n_k]$ decreases to 0 exponentially with EP $S_k$, see Eqs.~(\ref{EqSTspecialOne4},\ref{EqLimitCase1Uncertainty3},\ref{EqLimitCase1Uncertainty6}) and Figs.~\ref{FigCase1OneAbsorbTNVar} and \ref{FigCase3OneAbsorbTNVar}. On the contrary, for either $\{u_k\}$ tend to 0 or $\{w_k\}$ tend to infinity, both mean FPT $T_k$ and mean TNJ $n_k$ increase to infinity, while $\textrm{CV}[T_k]$ and $\textrm{CV}[n_k]$ are around 1, see Eqs.~(\ref{EqSTspecialOne2},\ref{EqSTspecialOne4},\ref{EqLimitCase2Uncertainty1},\ref{EqLimitCase2Uncertainty3}) and Figs.~\ref{FigCase2OneAbsorbTNVar} and \ref{FigCase4OneAbsorbTNVar}.

For $r\ll1$, uncertainty relations given in Eqs.~(\ref{EqLimitCase1Uncertainty3},\ref{EqLimitCase1Uncertainty5},\ref{EqLimitCase1Uncertainty6}) are established, while for For $r\gg1$, we have the one as given in Eq.~(\ref{EqLimitCase2Uncertainty3}). Numerical calculations show that, for randomly selected rates $\{u_k\}$ and $\{w_k\}$, the bounds given in Eqs.~(\ref{EqLimitCase1Uncertainty1},\ref{EqLimitCase1Uncertainty4},\ref{EqLimitCase1Uncertainty3},\ref{EqLimitCase1Uncertainty6}) are good approximations of $T_k, n_k, \textrm{CV}[T_k], \textrm{CV}[n_k]$ if EP $S_k$ are positive and large enough, see Fig.~\ref{FigRandominRateValuesUlarge}. Meanwhile, if EP $S_k$ is negative and its absolute value is large enough, the bounds given in Eqs.~(\ref{EqLimitCase2Uncertainty1},\ref{EqLimitCase2Uncertainty3}) are also good approximations of $T_k, n_k, \textrm{CV}[T_k], \textrm{CV}[n_k]$, see Fig.~\ref{FigRandominRateValuesUsmall}.

\subsection{Cases of equilibrium with $r=1$}\label{SectionRequal1}
%\textbf{(III)} Cases of equilibrium, {\it i.e.}, $r=1$ and $S_k=0$.
It can be shown directly from Eqs.~(\ref{EqSTsolutionOne},\ref{EqSnksolutionOne}), or by taking the limit $r\to1$ in Eqs.~(\ref{EqSTspecialOne2},\ref{EqSTspecialOne4}), that
\begin{eqnarray}
T_k&=&(N-k)(N+k+1)/2u,\label{EqTkEquilibrium}\\
n_k&=&(N-k)(N+k+1), \label{EqnkEquilibrium}
\end{eqnarray}
Both of them decrease with starting site $k$. Meanwhile, it can be obtained that, for $r=1$,
\begin{eqnarray}
&&\textrm{CV}[T_k]=\frac{(2N+1)^2+1}{3(N-k)(N+k+1)}-\frac23,\label{EqCVTkEquilibrium}\\
&&\textrm{CV}[n_k]=\frac{2N(N+1)+2k(k+1)-1}{3(N-k)(N+k+1)}=\textrm{CV}[T_k]-\frac{1}{(N-k)(N+k+1)},\label{EqCVnkEquilibrium}
\end{eqnarray}
both increase with starting site $k$. Numerical calculations show that Eqs.~(\ref{EqTkEquilibrium},\ref{EqnkEquilibrium},\ref{EqCVTkEquilibrium},\ref{EqCVnkEquilibrium}) are valid for general equilibrium cases, in which $w_k=u_k$ but may be different at different site $k$, see Fig.~\ref{FigCVTknkFiniteN}. Note that, for general equilibrium cases, $u$ in Eq.~(\ref{EqTkEquilibrium}) should be replaced by the harmonic mean of $\{u_i\}_{i=k}^{N-1}$.

\section{First passage process with infinite states}
For the first passage process depicted in Fig.~\ref{FigSchematic}\textbf{(b)}, with $-\infty$ a reflecting boundary, we obtain from Eqs.~(\ref{EqT},\ref{EqT2},\ref{Eqnk},\ref{Eqnk2}) that, for $u_k\equiv u$, $w_k\equiv w$ and $r=w/u$, %the formulation in Eq.~(\ref{EqSTspecialOne2}) can be simplified to
\begin{eqnarray}
&&T_k=\frac{N-k}{u-w},\quad n_k=\frac{(u+w)(N-k)}{u-w}=(u+w)T_k,\label{EqSTspecialOne2simplified1}\\
&&\textrm{CV}[T_k]=\frac{u+w}{(N-k)(u-w)}=\frac{1+\exp(-S_k/(N-k-1))}{(N-k)[1-\exp(-S_k/(N-k-1))]},\label{EqSTspecialOne2simplified2}\\
&&\textrm{CV}[n_k]=\frac{4uw}{(N-k)(u^2-w^2)}=\frac{4\exp(-S_k/(N-k-1))}{(N-k)[1-\exp(-2S_k/(N-k-1))]},
\label{EqSTspecialOne2simplified3}
\end{eqnarray}
where we assume $0\le r<1$, {\it i.e.}, $u>w$, otherwise $T_k=n_k=\infty$.

For fixed $w$ and $r, r_0\ll1$, $T_k(r)/T_k(r_0)\sim r/r_0=\exp(-\Delta S_k/(N-K-1))$, which is the same as listed in Eq.~(\ref{EqLimitCase1Relation}). Meanwhile, ${T_k^{2}}/{[T_k^{(2)}-T_k^{2}]}=1/\textrm{CV}[T_k]=n_k[(1-r)/(1+r)]^2\le n_k$, so the KUR is established, see Eq.~(\ref{EqLimitCase1Uncertainty5}) and \cite{Hiura2021}. From Eqs.~(\ref{EqSTspecialOne2simplified2},\ref{EqSTspecialOne2simplified3}), one can easily show $\textrm{CV}[n_k]/\textrm{CV}[T_k]={4uw}/{(u+w)^2}<1$, so the uncertainty of TNJ is always less than that of the FPT. Numerical calculations show that this is also true for the finite state cases provided $r<1$, see Fig.~\ref{FigCase1OneAbsorbTNVar},\ref{FigCase3OneAbsorbTNVar},\ref{FigCVnLessThanCVT}. Note that, the lower bounds listed in Eqs.~(\ref{EqLimitCase1Uncertainty3},\ref{EqLimitCase1Uncertainty6}) of $\textrm{CV}[T_k]$ and $\textrm{CV}[n_k]$ for finite state cases are the same as the ones given in Eq.~(\ref{EqSTspecialOne2simplified2},\ref{EqSTspecialOne2simplified3}) for infinite state cases. Meanwhile, Eq.~(\ref{EqCVnkEquilibrium}) and Fig.~\ref{FigCVTknkFiniteN} indicate that $\textrm{CV}[n_k]<\textrm{CV}[T_k]$ is also true for general equilibrium cases.

The processive motion of motor proteins (such as kinesin and dynein) along microtubule in cells can be well described by the one dimensional birth and death process depicted in Fig.~\ref{FigSchematic}\textbf{(b)} \cite{Howard2001,Kolomeisky2015,Mugnai2020}. Experiments show that the forward stepping of motor proteins is tightly coupled with ATP hydrolysis, {\it i.e.}, one ATP molecule is consumed in each forward step (about 8.2 nm), while the backward sliding is ATP free \cite{Schnitzer1997,Hua1997}. To analyze the energy efficiency of motor proteins, we need to know how many ATP molecules are consumed during the motion from starting site $k$ to reach boundary $N$ first.

Let $n_k^+$ ($n_k^-$) be the mean total numbers of forward (backward) jumps during the process. Obviously, $n_k^++n_k^-=n_k$ and $n_k^+-n_k^-=N-k$. So, according to Eq.~(\ref{EqSTspecialOne2simplified1}), we have $n_k^+=(N-k)/(1-r)$ and $n_k^-=r(N-k)/(1-r)$. For a given sample path of the first passage process with total number of forward jumps $\tilde{n}_k^+$ and total number of backward jumps $\tilde{n}_k^-$, the energy efficiency of ATP molecule is
$$
\tilde{\eta}^{\textrm{ATP}}_k=\frac{\tilde{n}_k^+-\tilde{n}_k^-}{\tilde{n}_k^+}=\frac{N-k}{\tilde{n}_k^+}.
$$
So $\langle1/\tilde{\eta}^{\textrm{ATP}}_k\rangle={\langle\tilde{n}_k^+\rangle}/{(N-k)}={n_k^+}/{(N-k)}=1/(1-r)$.
By Jensen's inequality, the mean value of energy efficiency $\eta^{\textrm{ATP}}_k=\langle\tilde{\eta}^{\textrm{ATP}}_k\rangle$ satisfies
\begin{eqnarray}
\eta^{\textrm{ATP}}_k\ge 1-r,\label{EqenergyEfficiency}
\end{eqnarray}
which decreases (increases) with ratio $r=w/u$ (EP $S_k$).

Moreover, it can be shown that CVs of total forward and backward jumps are as follows,
\begin{eqnarray}
\textrm{CV}[n_k^+]=\frac{r(1+r)}{(1-r)(N-k)},\quad \textrm{CV}[n_k^-]=\frac{1+r}{(1-r)r(N-k)}.\label{EqCVofnplusminus}
\end{eqnarray}
One can show that $\textrm{CV}[n_k^+]=r^2\textrm{CV}[n_k^-]$, and $\textrm{CV}[n_k^+]=(1+r)^2\textrm{CV}[n_k]/4<\textrm{CV}[n_k]=4r^2\textrm{CV}[n_k^-]/(1+r)^2<\textrm{CV}[n_k^-]$.
Evidently, $\textrm{CV}[n_k^+]$ increases (decreases) with ratio $r$ (EP $S_k$), while $\textrm{CV}[n_k^-]$ attains its minimum $(\sqrt{2}+1)^2/(N-k)$ when $r=\sqrt{2}-1$, at which $\eta^{\textrm{ATP}}_k\ge 2-\sqrt{2}$, $n_k^-=(N-k)/\sqrt{2}$, $n_k=(\sqrt{2}+1)(N-k)$ and $\textrm{CV}[n_k^+]=1/(N-k)$, $\textrm{CV}[n_k]=2/(N-k)$, $\textrm{CV}[T_k]=(\sqrt{2}+1)/(N-k)$. Although, both $n_k^+$ and $n_k^-$ decrease with starting site $k$, $\textrm{CV}[n_k^+]$ and $\textrm{CV}[n_k^-]$ increase with $k$.

\begin{figure}[htbp]
\includegraphics[width=0.8\linewidth]{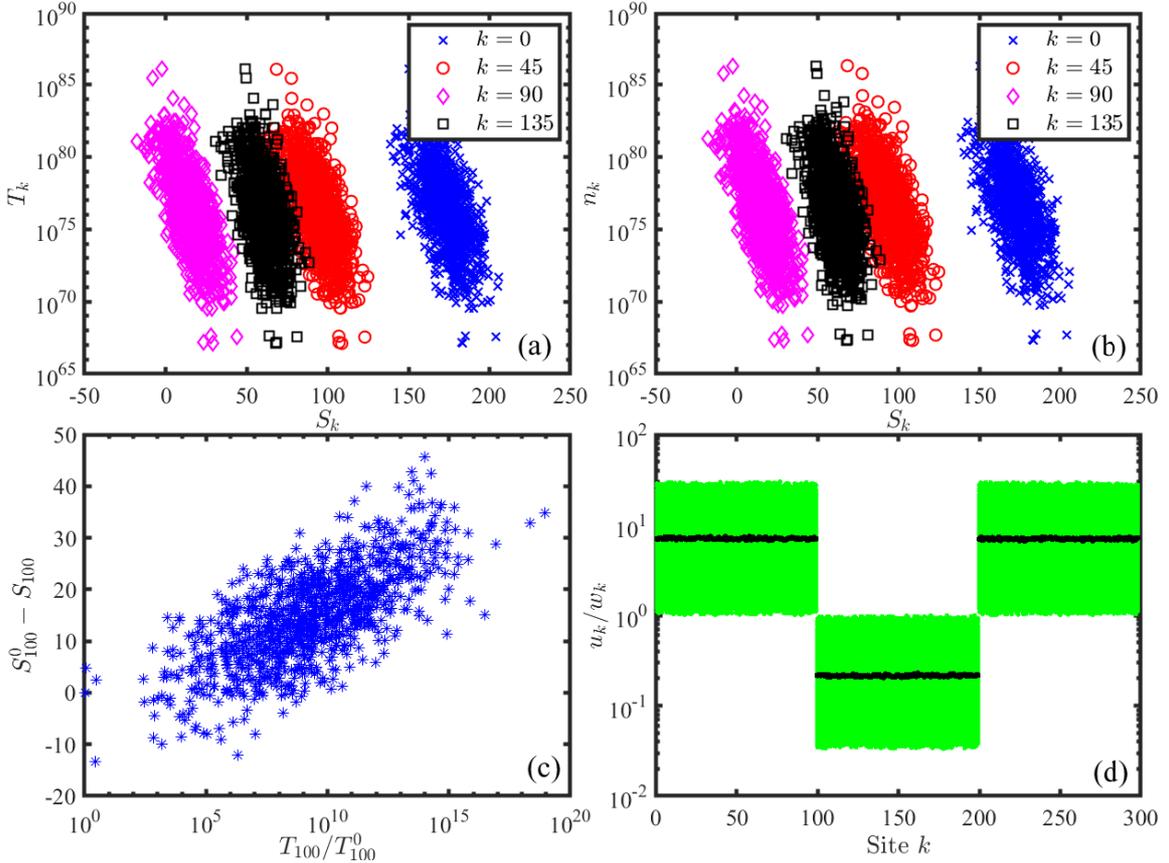}
\caption{Results for the first passage process as depicted in Fig.~\ref{FigSchematic}\textbf{(c)} (with $x_2$ therein assumed to be the absorbing boundary $N$ as in Fig.~\ref{FigSchematic}\textbf{(a)}). For $k<100$ and $k\ge200$, $\{u_k\}$ and $\{w_k\}$ are randomly selected in intervals $(0.5, 3)$ and $(0.1, 0.5)$ respectively, while for $100\le k<200$ their selection intervals are exchanged with each other. $N=300$ is used, and altogether 1000 groups of $\{u_k\}$ and $\{w_k\}$ are used. In \textbf{(d)}, ratio $u_k/w_k$ is plotted as small light dot, and their mean value is plotted as large thick dots. Values of $S_k$, $T_k$ and $n_k$ are obtained by formulations given in Eqs.~(\ref{EqSSsolutionOne},\ref{EqSTsolutionOne},\ref{EqSnksolutionOne}) respectively.}\label{FigTwoStateCasec}
\end{figure}

Finally, we address briefly the usual transition between two steady states, as depicted in Fig.~\ref{FigSchematic}\textbf{(c)}. With $x_2$ assumed to be the absorbing boundary $N$, the plots in Fig.~\ref{FigTwoStateCasec} show that, from any starting site $k$, both mean FPT $T_k$ and mean TNJ $n_k$ decrease with EP $S_k$, and roughly in an exponential manner. For the transition depicted in Fig.~\ref{FigSchematic}\textbf{(d)}, results are similar, see Fig.~\ref{FigTwoStateCased}.

\section{Conclusions}
In conclusion, the entropy production $S$ related properties of first passage process are discussed extensively in this study. For process with either finite or infinite internal states, approximated expressions of mean FPT $T$, mean TNJ $n$, and their coefficients of variation, $\textrm{CV}[T]$ and $\textrm{CV}[n]$, %as functions of EP $S$,
are obtained for either $S\gg0$ or $S\ll0$. Generally, $\textrm{CV}[n]<\textrm{CV}[T]$ for forward biased (or equilibrium) process, while for backward biased process, both of them tend to 1 quickly with the increase of the absolute value of entropy production $S$. Roughly speaking, mean FPT decreases exponentially with entropy production $S$. For the mean TNJ, it decreases exponentially first, and then tends to a starting site dependent limit finally. For equilibrium cases with $S=0$, approximated expressions of mean FPT $T$, mean TNJ $n$ and $\textrm{CV}[T]$, $\textrm{CV}[n]$ are also obtained, which mainly depend on the starting site and the position of absorbing boundary. Although most of expressions are obtained simply under the homogeneous assumption, in which values of forward rates and backward rates are independent of state, numerical calculations show they are also reasonable for general biased process, with forward rates always greater (or always less) than backward rates.

\clearpage
\newpage
\pagebreak
%\vspace*{1cm}

%\begin{appendix}

\setcounter{figure}{0}
\setcounter{section}{0}
\setcounter{equation}{0}
\setcounter{table}{0}
\setcounter{page}{1}
\renewcommand{\thefigure}{S\arabic{figure}}
\renewcommand{\thesection}{S\arabic{section}}
\renewcommand{\theequation}{S\arabic{equation}}
\renewcommand{\thetable}{S\Roman{table}}

\section*{\bf Supplemental Materials for \lq\lq entropy production related properties of first passage process"}\label{appendex}
\vspace*{0.3cm}
\centerline{Yunxin Zhang}

{\it Laboratory of Mathematics for Nonlinear Science, Shanghai Key Laboratory for Contemporary Applied Mathematics, Centre for Computational Systems Biology, School of Mathematical Sciences, Fudan University, Shanghai 200433, China.}

\vspace*{0.4cm}

The probability density $p_k(s,t)$ that a random particle initiated at site $k$ to reach the absorbing boundary $N$ at time $t$ and with EP $s$ satisfies the following backward master equation
\begin{equation}\label{EqSf}
\partial_t p_k(s,t)=u_kp_{k+1}(s-\ln(u_k/w_{k+1}),t)+w_kp_{k-1}(s-\ln(w_k/u_{k-1}),t)-(u_k+w_k)p_k(s,t),
\end{equation}
with $\ln(u_k/w_{k+1})$ and $\ln(w_k/u_{k-1})$ EPs ($k_B=1$) during one forward and backward jump, respectively \cite{Seifert2005,Seifert2012Stochastic,Teza2020}. From Eq.~(\ref{EqSf}), the mean EP $S_k:=\int_{-\infty}^{\infty}s\int_0^{\infty}p_k(s,t)\dt\ds$ during the passage process from site $k$ to boundary $N$ satisfies
\begin{equation}\label{EqSS}
u_kS_{k+1}+w_kS_{k-1}-(u_k+w_k)S_k=-\left(u_k\ln\frac{u_k}{w_{k+1}}+w_k\ln\frac{w_k}{u_{k-1}}\right).
\end{equation}
The mean FPT $T_k:=\int_0^{\infty}t\int_{-\infty}^{\infty}p_k(s,t)\ds\dt$, and the second moment of FPT $T_k^{(2)}:=\int_0^{\infty}t^2\int_{-\infty}^{\infty}p_k(s,t)\ds\dt$, satisfy
\begin{eqnarray}
&&u_kT_{k+1}+w_kT_{k-1}-(u_k+w_k)T_k=-1,\label{EqST}\\
&&u_kT_{k+1}^{(2)}+w_kT_{k-1}^{(2)}-(u_k+w_k)T_k^{(2)}=-2T_k.\label{EqST2}
\end{eqnarray}

Meanwhile, the probability density $q_k(n)$ that a random particle initiated at site $k$ to reach the absorbing boundary $N$ with TNJ $n$ satisfies the following backward master equation
\begin{equation}\label{EqSq}
u_kq_{k+1}(n-1)+w_kq_{k-1}(n-1)-(u_k+w_k)q_k(n)=0.
\end{equation}
From which, the mean TNJ $n_k:=\sum_{n=0}^{\infty}nq_k(n)$, and the corresponding second moment $n_k^{(2)}:=\sum_{n=0}^{\infty}n^2q_k(n)$,
satisfy
\begin{eqnarray}
&&u_kn_{k+1}+w_kn_{k-1}-(u_k+w_k)n_k=-(u_k+w_k),\label{EqSnk}\\
&&u_kn_{k+1}^{(2)}+w_kn_{k-1}^{(2)}-(u_k+w_k)n_k^{(2)}=-(2n_k-1)(u_k+w_k).\label{EqSnk2}
\end{eqnarray}

By usual algebraic calculation, one can obtain from Eq.~(\ref{EqSS}) that
\begin{eqnarray}
S_k&=&\sum_{l=k}^{N-2}\ln\frac{u_l}{w_{l+1}}=\ln\left(\prod_{l=k}^{N-2}\frac{u_l}{w_{l+1}}\right), \label{EqSSsolutionOne}
\end{eqnarray}
the solution $T_k$ of Eq.~(\ref{EqST}) is
\begin{equation}\label{EqSTsolutionOne}
T_k=\sum_{l=k}^{N-1}\sum_{i=0}^l\left[\frac{1}{u_i}\prod_{j=i+1}^l\left(\frac{w_j}{u_j}\right)\right],
\end{equation}
and according to Eq.~(\ref{EqST2},\ref{EqSnk},\ref{EqSnk2}),
\begin{eqnarray}
T_k^{(2)}%^{\rm (I)}
&=&\sum_{l=k}^{N-1}\sum_{i=0}^l\left[\frac{2T_i}{u_i}\prod_{j=i+1}^l\left(\frac{w_j}{u_j}\right)\right],\label{EqST2solutionOne}\\
n_k%^{\rm (I)}
&=&\sum_{l=k}^{N-1}\sum_{i=0}^l\left[\left(1+\frac{w_i}{u_i}\right)\prod_{j=i+1}^l\left(\frac{w_j}{u_j}\right)\right],\label{EqSnksolutionOne}\\
n_k^{(2)}&=&\sum_{l=k}^{N-1}\sum_{i=0}^l\left[(2n_i-1)\left(1+\frac{w_i}{u_i}\right)\prod_{j=i+1}^l\left(\frac{w_j}{u_j}\right)\right].\label{EqSnk2solutionOne}
\end{eqnarray}

\begin{figure}[htbp]
\includegraphics[width=0.8\linewidth]{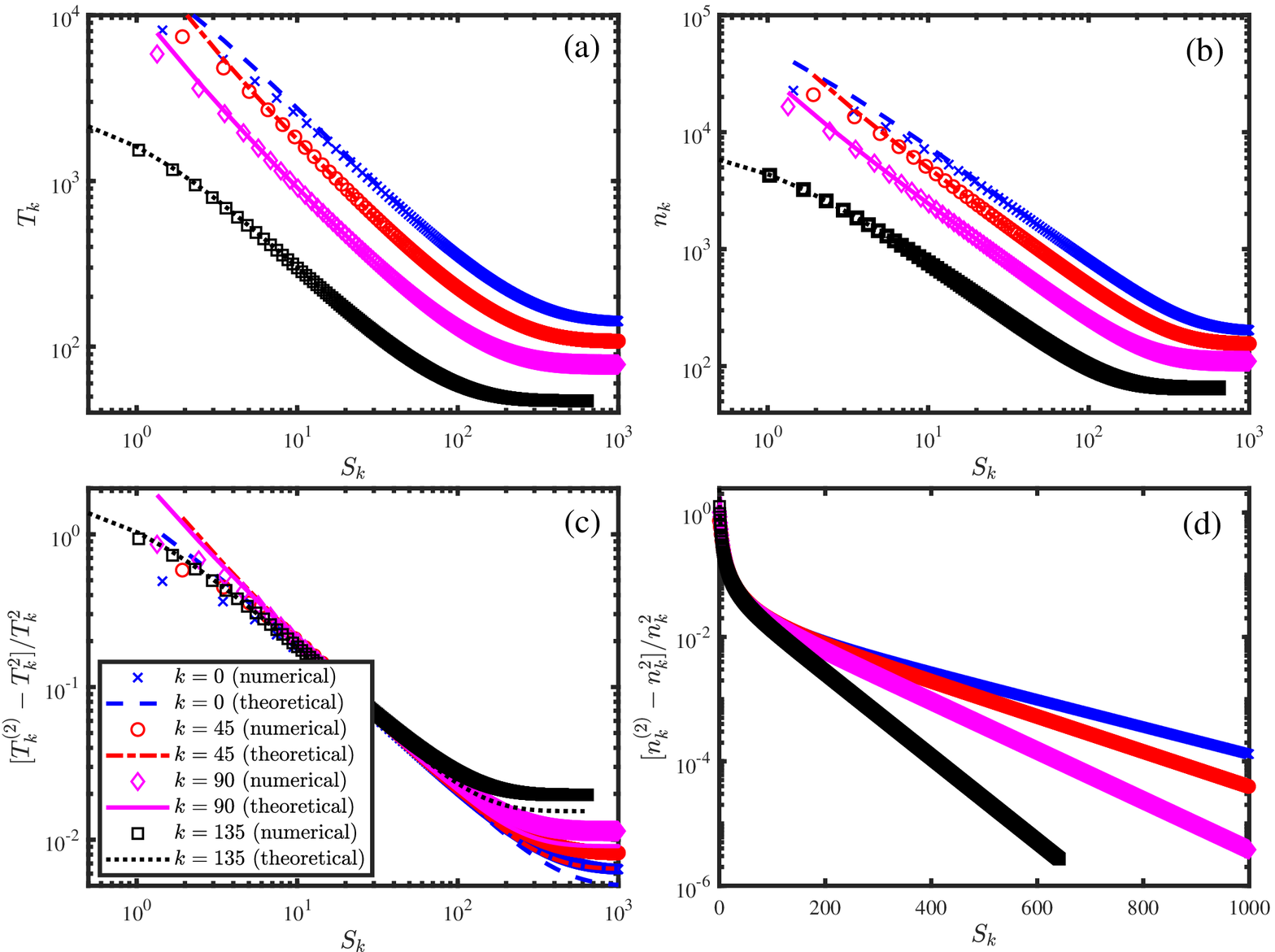}
	\caption{Similar plots as in Fig.~\ref{FigCase1OneAbsorbTNVar}. In the first calculation, all forward rates $\{u_k^{(1)}\}$ and backward rates $\{w_k^{(1)}\}$ are randomly selected in interval $(0.5, 3)$. Then in following calculations, forward rates $\{u_k\}$ remain unchanged while backward rates change according to $w_k^{(l+1)}=\gamma w_k^{(l)}$ with $\gamma=0.99$.
Points are obtained numerically by formulations given in Eqs.~(\ref{EqSSsolutionOne},\ref{EqSTsolutionOne},\ref{EqST2solutionOne},\ref{EqSnksolutionOne},\ref{EqSnk2solutionOne}).
The theoretical line for $T_k$ is obtained by ${(N-k)}/{[u(1-r)]}$  , and other lines
are obtained from the bounds given in Eqs.~(\ref{EqLimitCase1Uncertainty4},\ref{EqLimitCase1Uncertainty3},\ref{EqLimitCase1Uncertainty6}), respectively. Again, $r$ is replaced by $\exp(-S_k/(N-k-1))$, and $u$ is replaced by the harmonic mean of $\{u_i\}_{i=k}^{N-1}$.
}\label{FigCase3OneAbsorbTNVar}
\end{figure}

\begin{figure}[htbp]
\includegraphics[width=0.5\linewidth]{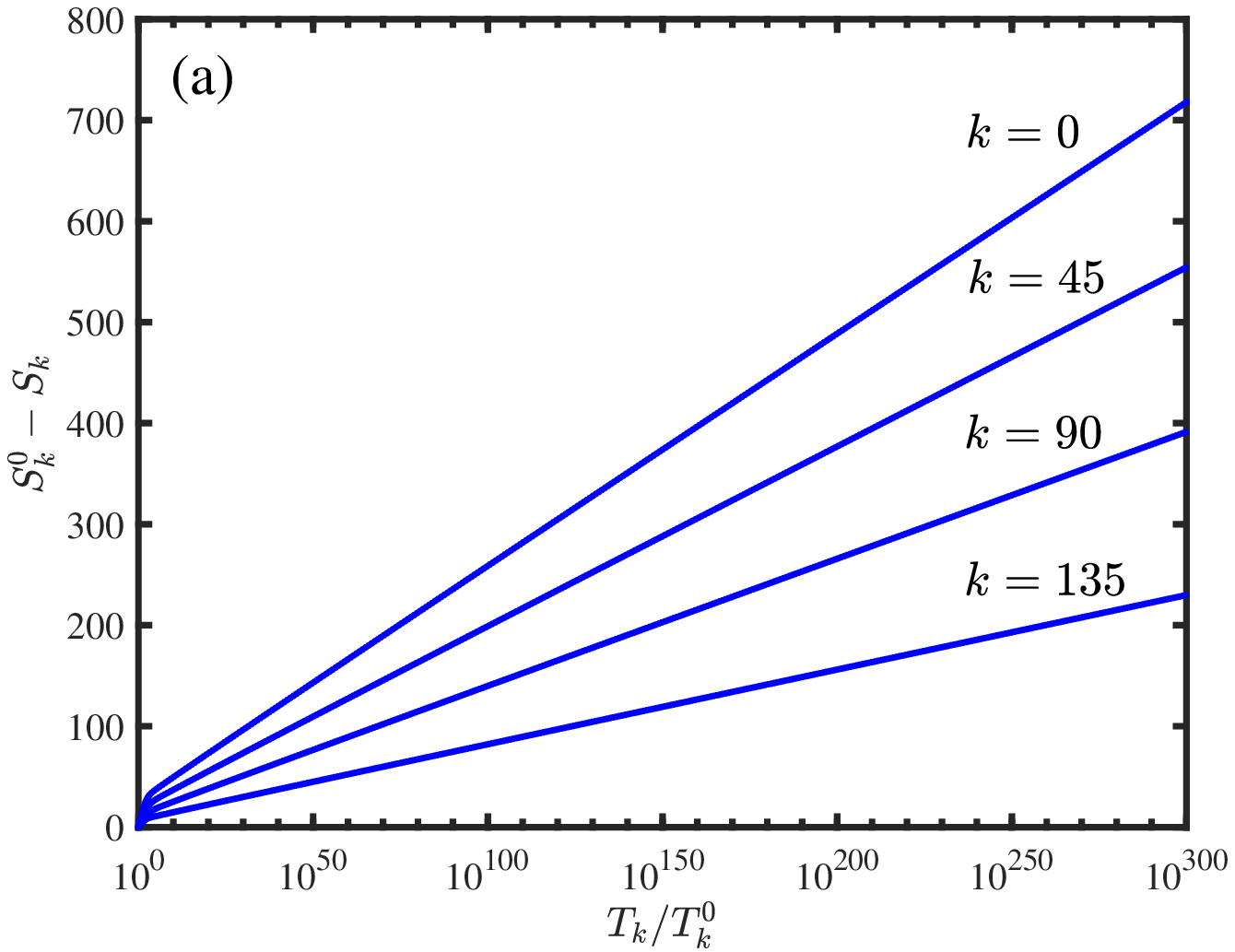}\includegraphics[width=0.5\linewidth]{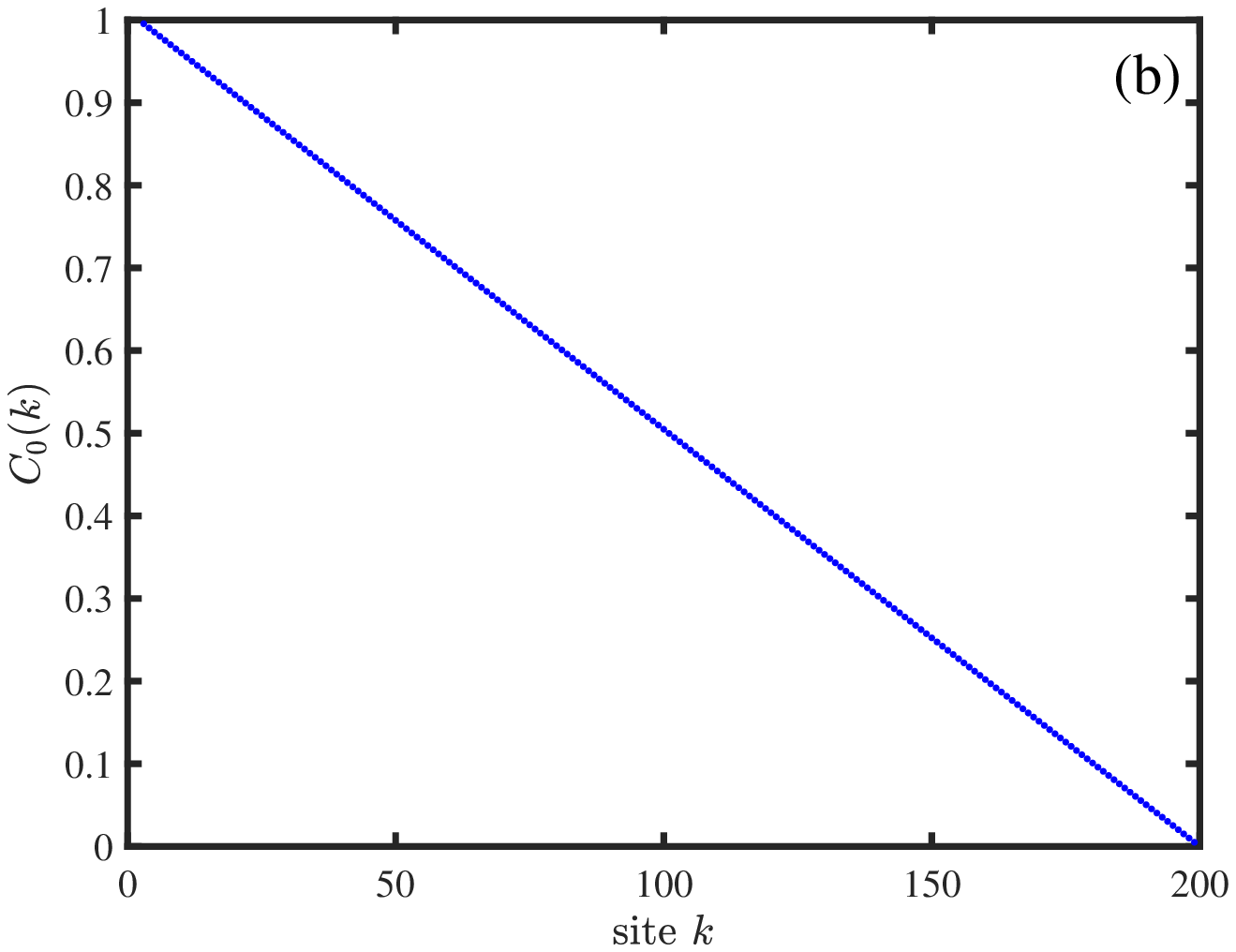}
	\caption{Similar as in Fig.~\ref{FigCase1OneAbsorb} but with $\gamma=u_k^{(l+1)}/u_k^{(l)}\equiv0.99$, which means forward rates $\{u_k\}$ tend to zero. \textbf{(a)} $S^0_{k}-S_{k}$ versus $T_{k}/T^0_{k}$ for $k=0,45,90,135$, and \textbf{(b)} $C_0(k)$ obtained by fitting $T_k/T_k^0=C_1(k)\exp(-(S_k-S_k^0)/C_0(k))$, see Eq.~(\ref{EqLimitCase2Relation-1}).
For $\{u_k\}$ fixed while $\{w_k\}$ increase to infinity, figures \textbf{(a,b)} are almost unchanged,
see Eqs. (\ref{EqLimitCase2Relation-1},\ref{EqLimitCase4Relation-1}).
Values of $S_k$ and $T_k$ are obtained  by Eqs.~(\ref{EqSSsolutionOne},\ref{EqSTsolutionOne}), respectively.}
	\label{FigCase2OneAbsorb}
\end{figure}

\begin{figure}[htbp]
\includegraphics[width=0.8\linewidth]{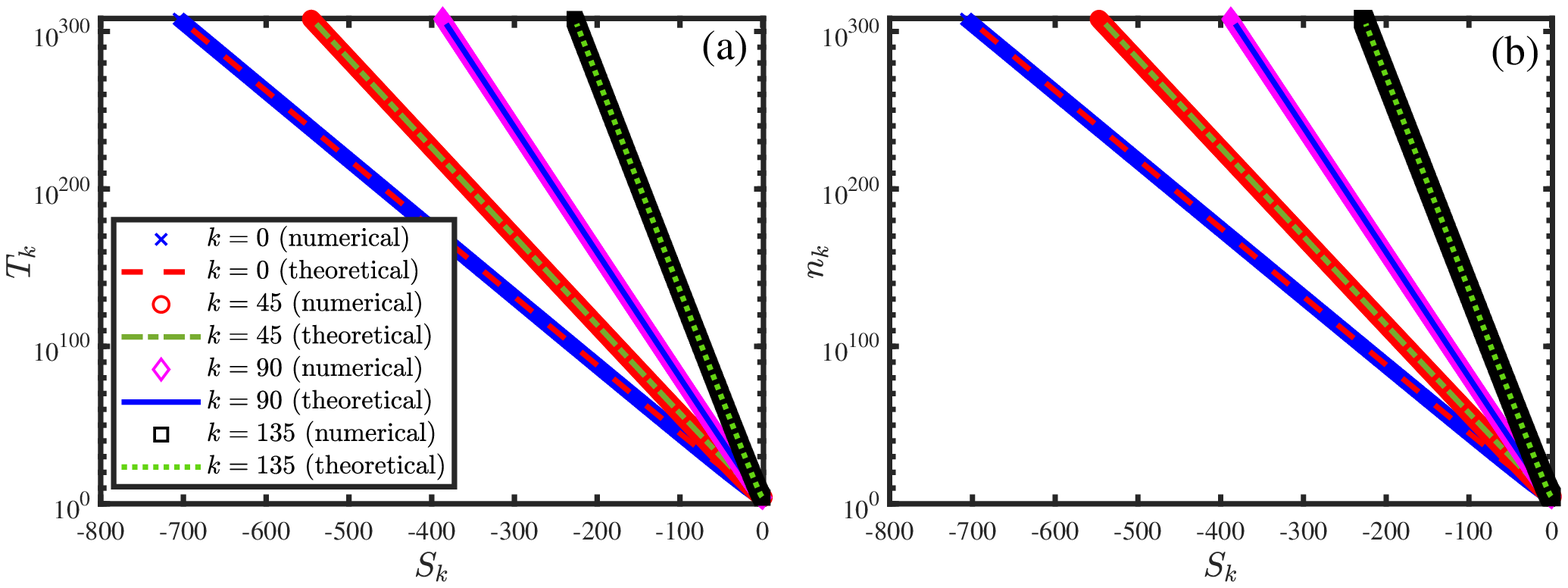}
	\caption{Similar plots as in Fig.~\ref{FigCase1OneAbsorbTNVar}, but with parameter values chosen by the same method as in Fig.~\ref{FigCase2OneAbsorb}. For these cases, $[T_k^{(2)}-T_k^{2}]/T_k^2$ and $[n_k^{(2)}-n_k^{2}]/n_k^2$ are almost constant 1 for $S_k\le10$, see Eq.~(\ref{EqLimitCase2Uncertainty3}).}\label{FigCase2OneAbsorbTNVar}
\end{figure}

\begin{figure}[htbp]
\includegraphics[width=0.8\linewidth]{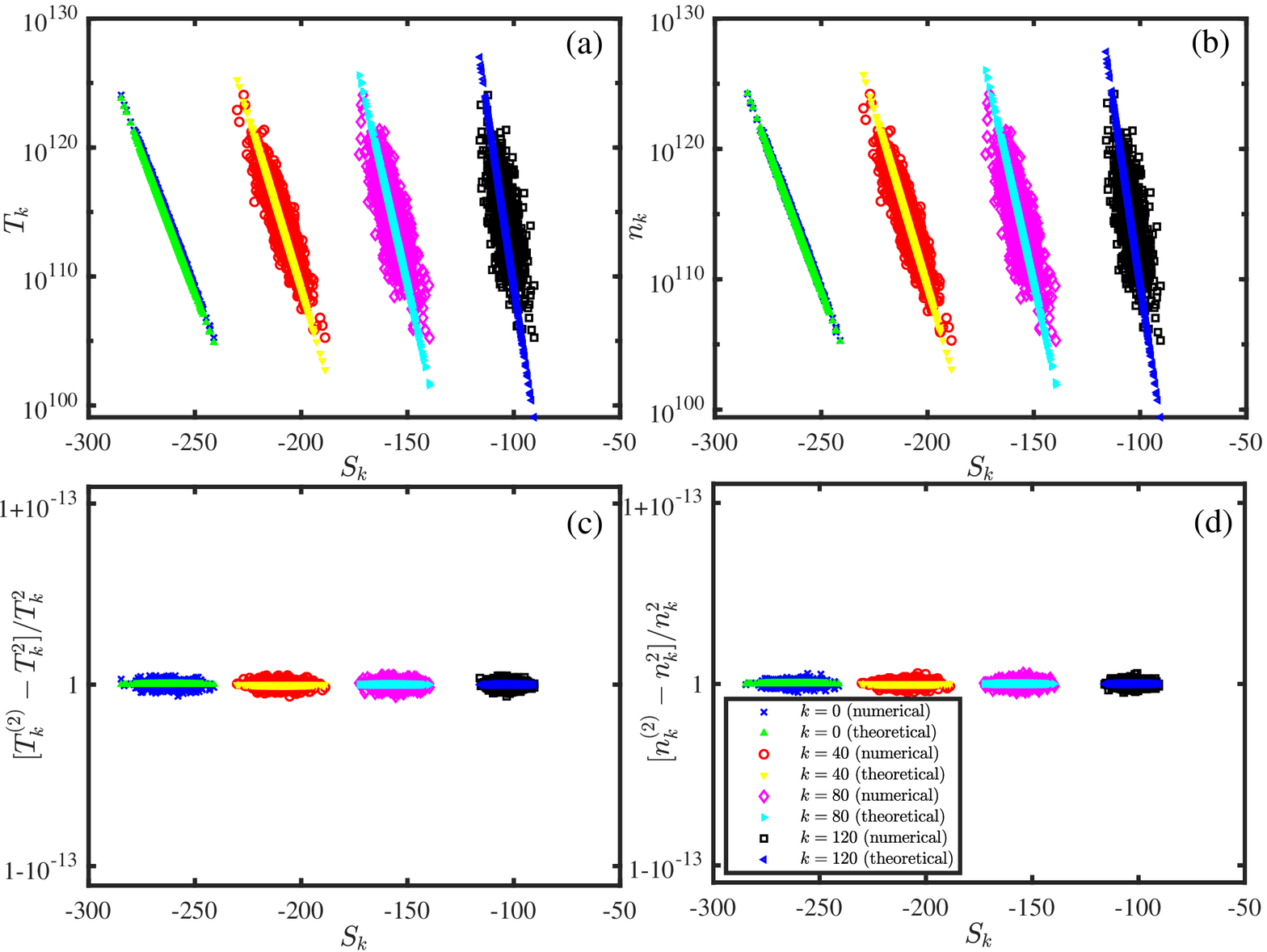}
	\caption{Similar plots as Fig.~\ref{FigRandominRateValuesUlarge}, but with $\{u_i\}$ are randomly selected in interval $(0.5,1)$. The theoretical lines (marked by varies triangles) are obtained from the bounds given in Eqs.~(\ref{EqLimitCase2Uncertainty1},\ref{EqLimitCase2Uncertainty3}), with $r$ replaced by $\exp(-S_k/(N-k-1))$, and $w$ replaced by the harmonic mean of $\{w_i\}_{i=k}^{N-1}$. Numerical values are obtained by formulations given in Eqs.~(\ref{EqSSsolutionOne},\ref{EqSTsolutionOne},\ref{EqST2solutionOne},\ref{EqSnksolutionOne},\ref{EqSnk2solutionOne}). }\label{FigRandominRateValuesUsmall}
\end{figure}

\begin{figure}[htbp]
\includegraphics[width=0.8\linewidth]{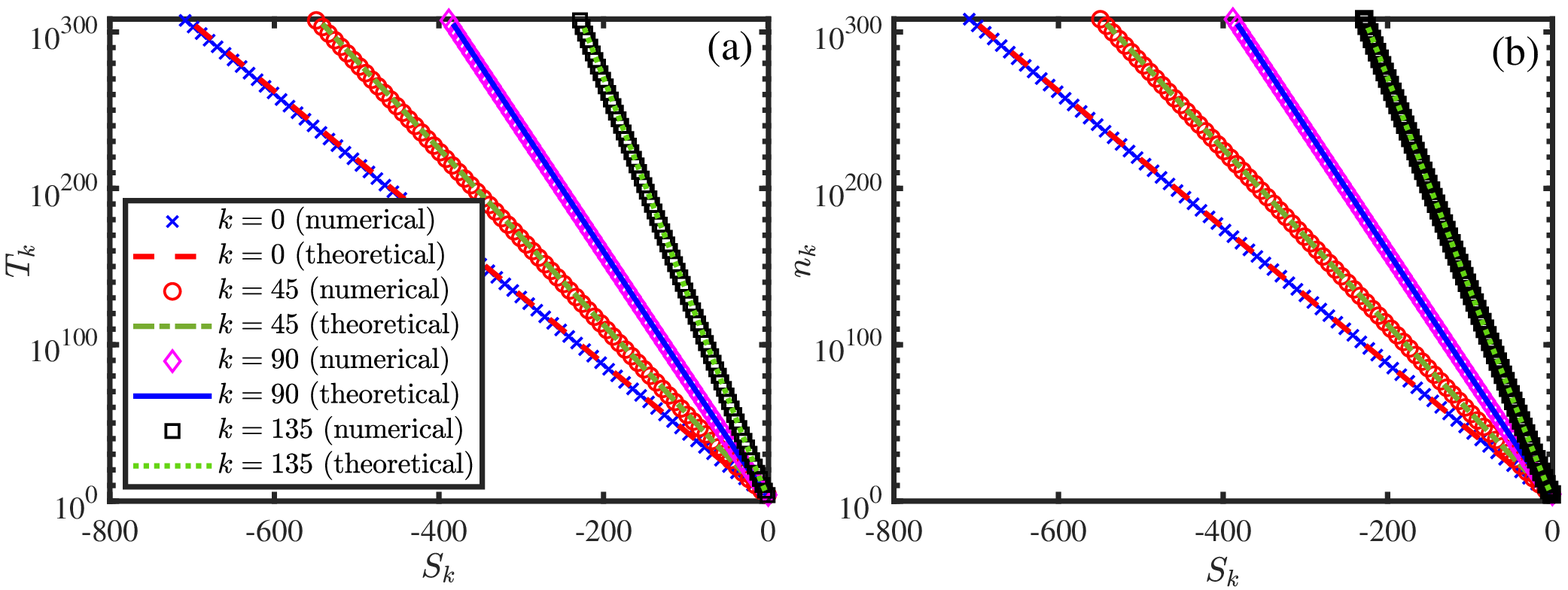}
	\caption{Similar plots as in Fig.~\ref{FigCase3OneAbsorbTNVar}, but $w_k^{(l+1)}=\gamma w_k^{(l)}$ with $\gamma=1.05$. }\label{FigCase4OneAbsorbTNVar}
\end{figure}

\begin{figure}[htbp]
\includegraphics[width=0.8\linewidth]{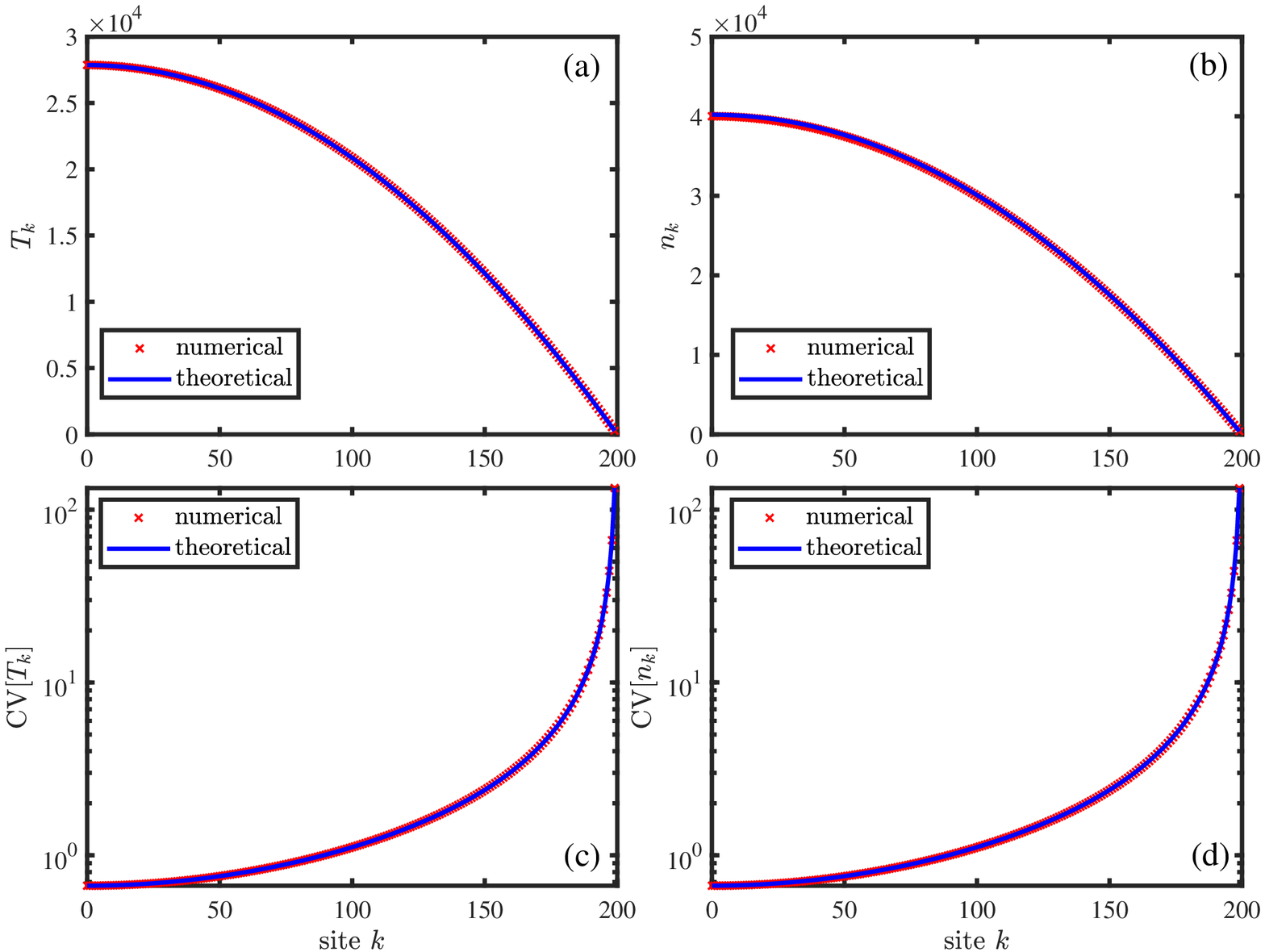}
	\caption{Mean FPT $T_k$, mean TNJ $n_k$ and their coefficients of variation $\textrm{CV}[T_k]$, $\textrm{CV}[n_k]$ for general equilibrium cases. In each calculations, $\{w_k\}$ are randomly selected in interval $(0.5,1)$, and then let $u_k=w_k$. The numerical results are obtained by the average of 1000 calculations (see Eqs.~(\ref{EqSSsolutionOne},\ref{EqSTsolutionOne},\ref{EqST2solutionOne},\ref{EqSnksolutionOne},\ref{EqSnk2solutionOne}) for formulations), and the theoretical ones are obtained according to Eqs.~(\ref{EqTkEquilibrium},\ref{EqnkEquilibrium},\ref{EqCVTkEquilibrium},\ref{EqCVnkEquilibrium}), which are derived for the special cases with $u_k\equiv u=w\equiv w_i$ for any $k,i$. }\label{FigCVTknkFiniteN}
\end{figure}

\begin{figure}[htbp]
\includegraphics[width=0.8\linewidth]{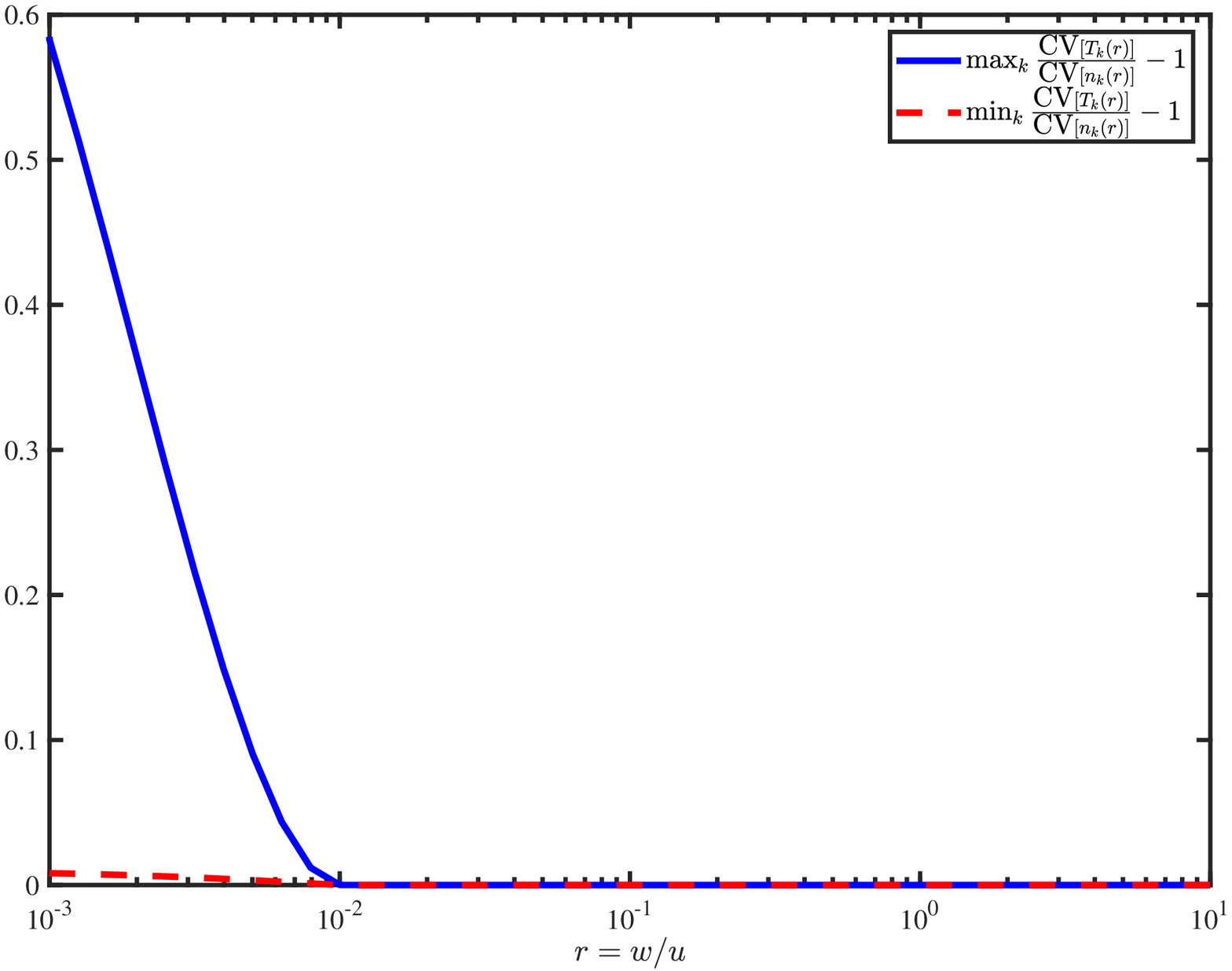}
	\caption{The maximum and minimum of $\textrm{CV}[T_k(r)]/\textrm{CV}[n_k(r)]-1$, for $N=100$ and $r=w/u$ changes from $10^{-3}$ to $10$. This plot shows that $\textrm{CV}[n_k(r)]\le \textrm{CV}[T_k(r)]$ holds true for the case with finite states as depicted in Fig.~\ref{FigSchematic}{\bf (a)}, and $\textrm{CV}[n_k(r)]\sim \textrm{CV}[T_k(r)]$ for large $r$ as obtained in Eq.~(\ref{EqLimitCase2Uncertainty3}).}\label{FigCVnLessThanCVT}
\end{figure}

\begin{figure}[htbp]
\includegraphics[width=0.8\linewidth]{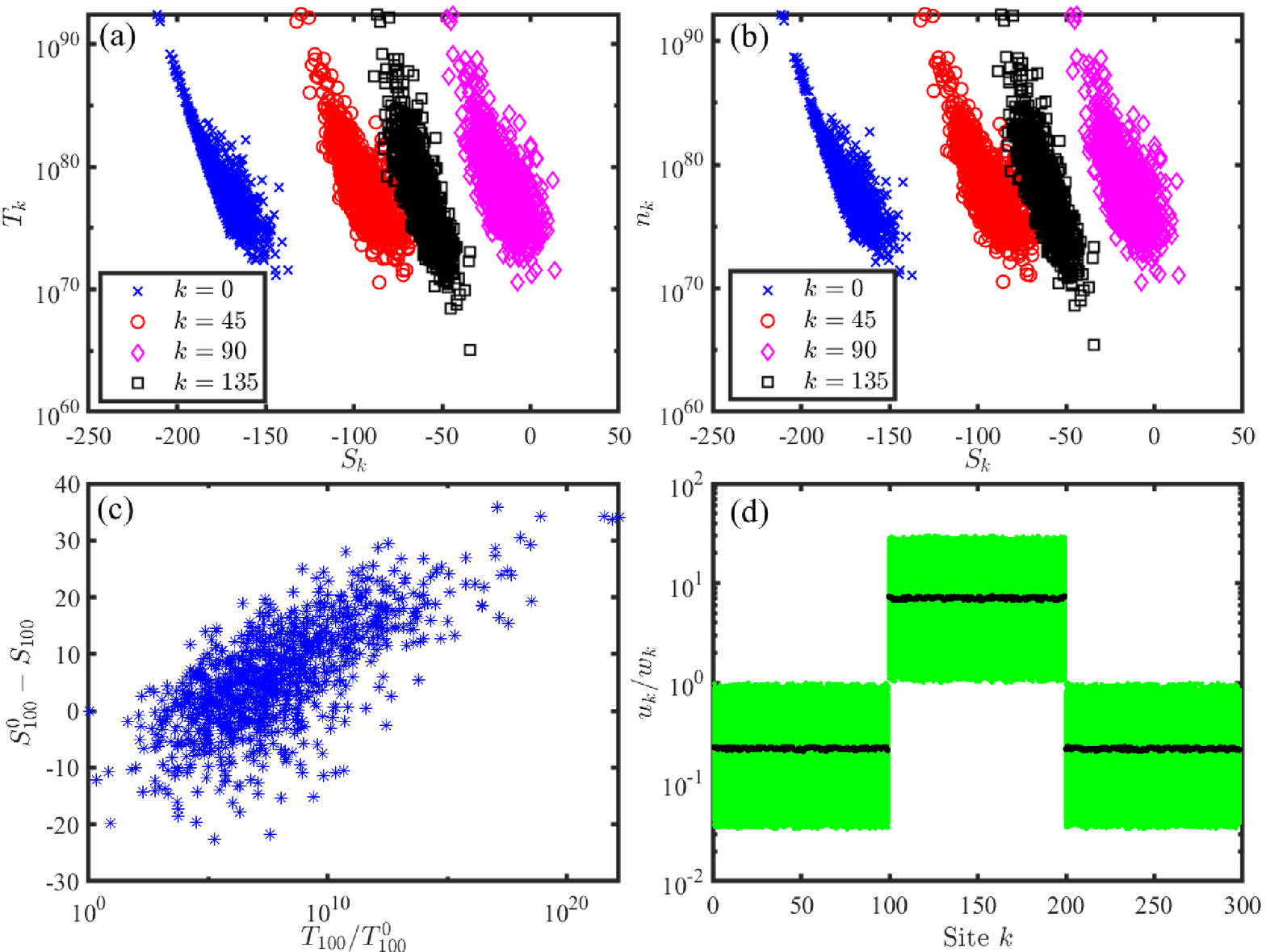}
\caption{Results for the first passage process as depicted in Fig.~\ref{FigSchematic}\textbf{(d)} (with $x_2$ assumed to be the absorbing boundary $N$ as in Fig.~\ref{FigSchematic}\textbf{(a)}). For $k<100$ and $k\ge200$, $\{u_k\}$ and $\{w_k\}$ are randomly selected in intervals $(0.1, 0.5)$ and $(0.5, 3)$ respectively, while for $100\le k<200$ their selection intervals are exchanged with each other. The same as in Fig.~\ref{FigTwoStateCasec}, $N=300$ is used, and altogether 1000 groups of $\{u_k\}$ and $\{w_k\}$ are used. In \textbf{(d)}, large thick dots represent the mean value of $u_k/w_k$, and the 1000 values of $u_k/w_k$ are plotted as small light dots.}	\label{FigTwoStateCased}
\end{figure}

\end{document}